\documentclass[aps,prl,reprint,superscriptaddress,floatfix,showkeys]{revtex4-1}

\usepackage{graphicx}
\usepackage{dcolumn}
\usepackage{bm}

\begin{document}


\title{$\bm{^{14}\textrm{Be}(p,n)^{14}\textrm{B}}$ reaction at 69 MeV 
in inverse kinematics}


\author{Y.~Satou}
\email[]{satou@snu.ac.kr}
\affiliation{Department of Physics and Astronomy, 
Seoul National University, 599 Gwanak, Seoul 151-742, Republic of Korea}
\author{T.~Nakamura}
\author{Y.~Kondo}
\author{N.~Matsui}
\author{Y.~Hashimoto}
\author{T.~Nakabayashi}
\author{T.~Okumura}
\author{M.~Shinohara}
\affiliation{Department of Physics, 
Tokyo Institute of Technology, 2-12-1 Oh-Okayama, Meguro, Tokyo 152-8551, Japan}
\author{N.~Fukuda}
\author{T.~Sugimoto}
\author{H.~Otsu}
\author{Y.~Togano}
\author{T.~Motobayashi}
\author{H.~Sakurai}
\author{Y.~Yanagisawa}
\author{N.~Aoi}
\author{S.~Takeuchi}
\author{T.~Gomi}
\author{M.~Ishihara}
\affiliation{RIKEN Nishina Center, 2-1 Hirosawa, Wako, Saitama 351-0198, Japan}
\author{S.~Kawai}
\affiliation{Department of Physics, Rikkyo University, 
3 Nishi-Ikebukuro, Toshima, Tokyo 171-8501, Japan}
\author{H.J.~Ong}
\author{T.K.~Onishi}
\affiliation{Department of Physics, University of Tokyo, 
7-3-1 Hongo, Bunkyo, Tokyo 113-0033, Japan}
\author{S.~Shimoura}
\author{M.~Tamaki}
\affiliation{Center for Nuclear Study (CNS), University of Tokyo, 
2-1 Hirosawa, Wako, Saitama 351-0198, Japan}
\author{T.~Kobayashi}
\author{Y.~Matsuda}
\author{N.~Endo}
\author{M.~Kitayama}
\affiliation{Department of Physics, Tohoku University, 
Aoba, Sendai, Miyagi 980-8578, Japan}


\date{\today}

\begin{abstract}
A Gamow-Teller (GT) transition from the drip-line nucleus $^{14}\textrm{Be}$ 
to $^{14}\textrm{B}$ 
was studied via the $(p,n)$ reaction in inverse kinematics 
using a secondary $^{14}\textrm{Be}$ beam at 69 MeV/nucleon. 
The invariant mass method is employed 
to reconstruct the energy spectrum. 
A peak is observed at an excitation energy of 1.27(2) MeV in $^{14}\textrm{B}$, 
together with bumps at 2.08 and 4.06(5) MeV. 
The observed forward peaking of the state at 1.27 MeV 
and a good description for the differential cross section, 
obtained 
with a DWBA calculation provide support for the $1^+$ assignment to this state. 
By extrapolating the cross section to zero momentum transfer 
the GT-transition strength is deduced. 
The value is found to compare well with that reported in 
a $\beta$-delayed neutron emission study. 
\end{abstract}

\keywords{Gamow-Teller transition, 
$^1\textrm{H}$$(^{14}\textrm{Be},$$^{13}\textrm{B}$+$n)$, 
Unbound states, 
Distorted-wave approximation}

\maketitle
The Gamow-Teller (GT) transition, 
characterized by an angular momentum transfer of $\Delta L$=0, 
a spin transfer of $\Delta S$=1, 
and an isospin transfer of $\Delta T$=1 between initial and final nuclei, 
represents one of the fundamental modes of nuclear 
excitations. 
Its properties are important not only for understanding 
spin-isospin correlations in nuclei~\cite{Harakeh01}, 
but also for their implications for a variety of astrophysical phenomena 
where weak processes play a key role, 
such as late stages of stellar evolution 
and neutrino nucleosynthesis~\cite{Langanke03}. 
The rapid increase in the availability of Radioactive Isotope (RI) beams 
in recent years 
has made it possible to study GT strengths 
in nuclei far from the $\beta$-stability line. 
Although $\beta$ decay has been the major source of information 
about such strengths, 
it might be expected that nuclear charge-exchange (CE) reactions 
play an important role~\cite{Austin97}, 
with their distinguishing feature 
of being unconstrained by any $Q_{\beta}$ limitations. 
We performed an experimental study of the $(p,n)$ reaction 
on a neutron-rich nucleus $^{14}\textrm{Be}$ 
using the invariant mass method in inverse kinematics. 
The data on the $^{14}\textrm{Be}(\textrm{g.s.})$
$\rightarrow$$^{14}\textrm{B}(1^+_1)$ 
transition process enabled the extraction of $B(\textrm{GT})$ 
from forward angle $(p,n)$ cross sections. 
Studies were also made on states populated 
in the $^1\textrm{H}$$(^{14}\textrm{Be},$$^{13}\textrm{B}$+$n)$ reaction. 

The $^{14}\textrm{Be}$ nucleus is located at the neutron drip line. 
A two-neutron halo structure
has been suggested by inclusive~\cite{Tanihata88,Zahar93} 
as well as exclusive~\cite{Labiche01} reaction studies. 
Unambiguous identification of the first $2^+$ state 
was made recently~\cite{Sugimoto07}. 
The $\beta$ decay of $^{14}\textrm{Be}$ 
is known to be dominated by one-neutron emission 
with a branching ratio compatible with 100\%~\cite{Bergmann99}. 
A concentration of the decay strength in a state at 1.28 MeV, 
which is located just above the one-neutron separation energy 
of $S_n$=0.97(2) MeV~\cite{Audi03} in $^{14}\textrm{B}$, 
was found in energy spectra of decay neutrons~\cite{Belbot97,Aoi02}. 
The small $\log ft$ value of 3.68(5)~\cite{Aoi02} 
has indicated that the decay is due to a GT transition, 
providing the $1^+$ assignment to this state. 

Existing CE studies on $\beta$-unstable nuclei 
have been based on two major methods: 
the first involves final states in a bound region 
and the missing mass technique was applied, 
where momenta of heavy charged ejectile were measured 
either by using a spectrograph 
in the $^6\textrm{He}(p,n)^6\textrm{Li}$~\cite{Brown96,Cortina-Gil96} and 
$^{34}\textrm{P}(^7\textrm{Li},$$^7\textrm{Be}$+
$\gamma)$$^{34}\textrm{Si}$~\cite{Zegers10} reactions, 
or a telescope in the $^6\textrm{He}(p,n)^6\textrm{Li}$ reaction~\cite{Li02}. 
The second method involves unbound final states, 
and the invariant mass technique 
was adopted. 
All particles from a decay of an unbound state were detected 
using a plastic counter hodoscope 
in the $^{11}\textrm{Li}(p,n)$$^{11}\textrm{Be}$~\cite{Teranishi97} 
and $^{14}\textrm{Be}(p,n)$$^{14}\textrm{B}$~\cite{Takeuchi01} reactions; 
isobaric analog states in the residual nuclei 
were identified. 
Until now, 
there exist only a limited number of CE studies 
involving unstable nuclear beams; 
it is desirable to further explore experimental approaches 
which permit reliable determination 
of relevant nuclear structure information. 
The present study, 
succeeding to the latter works, 
represents the first extraction of the GT strength 
on an unstable nucleus, 
leading to a particle unbound state, in an inverse kinematics CE reaction. 
Independent detectors 
were used for different decay products, charged fragment and neutron, 
to achieve high enough acceptance near the threshold, 
together with a liquid hydrogen target to increase the reaction yield. 
In inverse kinematics, 
decay neutrons emitted from the unbound system, 
which is Lorentz boosted with a beam-like velocity in the laboratory, 
acquire high energies and can be detected with high efficiencies. 
With this new feature, it is shown that the inverse kinematics CE reaction 
accompanied by neutron decay 
provides an alternative approach for nuclear structure, 
to $\beta$-delayed neutron spectroscopy. 

The extraction of the GT strength in CE reactions 
relies on the proportionality relationship between 
the reduced transition probability $B(\textrm{GT})$ and the cross section 
at zero momentum transfer ($q$=0)~\cite{Goodman80,Taddeucci87}: 
\begin{equation}
(d\sigma/d\Omega)_{q=0}^{L=0}
=\left[\frac{E_iE_f}{\left(\hbar^2c^2\pi\right)^2}\right]
N_D|J_{\sigma \tau}|^2B(\textrm{GT}). 
\label{eq:proportionality_1}
\end{equation}
Here $E_{i(f)}$ is the reduced energy in the initial (final) channel 
and $|J_{\sigma \tau}|$ is the volume integral 
of the central part of the effective nucleon-nucleon ($NN$) interaction 
in the spin-isospin channel. 
The distortion factor 
$N_D$=$(d\sigma/d\Omega)_{q=0}^{\rm DW}/(d\sigma/d\Omega)_{q=0}^{\rm PW}$ 
is the ratio of distorted-wave (DW) and plane-wave (PW) cross sections. 
The cross section $(d\sigma/d\Omega)_{q=0}^{L=0}$, 
defining the $L$=0 contribution at $q$=0, 
is obtained by extrapolating the 0$^{\circ}$ cross section to $q$=0 
using the ratio of distorted-wave Born approximation (DWBA) cross sections 
at $q$=0 and 0$^{\circ}$, 
with an additional factor given by the ratio of the cross section 
calculated with only the central terms of the $NN$ interaction and 
that with the full force components. 
Although the above relationship [Eq.(\ref{eq:proportionality_1})] has been 
mainly used for CE reaction data at bombarding energies above 100 MeV/nucleon, 
it is expected to be applicable for $(p,n)$ scattering data 
involving transitions with appreciable strengths, 
at energies of the present work (around $E_p$=70 MeV), 
due to the simple reaction mechanism of the nucleon scattering and 
to the availability of modern nucleon-nucleus optical model potentials (OMPs). 

The experiment was performed 
at the projectile-fragment separator RIPS~\cite{Kubo92}, 
which is a part of the RIBF accelerator complex 
operated by RIKEN Nishina Center. 
The setup was identical to the one used in Ref.~\cite{Kondo10}. 
The radioactive beam of $^{14}\textrm{Be}$ 
was produced from a 100-MeV/nucleon $^{18}\textrm{O}$ beam, 
which impinged on a 6-mm-thick Be target. 
The typical beam intensity 
was 7 kcps with a momentum spread of $\Delta P/P$=$\pm 2\%$. 
The beam profile was monitored by 
a pair of parallel-plate avalanche counters (PPACs), 
placed upstream of the target. 
The secondary target was a liquid hydrogen target~\cite{Ryuto05} 
with a cylindrical shape: 
3 cm in diameter and 229$\pm$6 mg/cm$^2$ in thickness. 
The average energy of $^{14}\textrm{Be}$ at the target center 
was 69 MeV/nucleon. 
The target was surrounded by a NaI(Tl) scintillator array 
to detect $\gamma$ rays from charged fragments. 
The fragments bent by a dipole magnet behind the target 
were detected by a plastic counter hodoscope, 
which gave $Z$ identification by $\Delta E$. 
Their trajectory was measured by a set of multi-wire drift chambers 
placed before and after the magnet, 
which, together with time of flight, 
gave mass identification. 
Neutrons were detected by two walls 
of plastic scintillator arrays placed 4.6 and 5.8 m downstream of the target. 
The neutron detection efficiency 
was $24.1\pm0.8\%$ for a threshold setting of 4 MeVee. 
The relative energy $E_{\rm rel}$ of the final system was calculated 
from the momentum vectors of the charged fragment and the neutron. 
The bending magnet for the fragment allowed the two detector systems 
to be located separately. 
This, together with the kinematical focusing of the reaction products toward 
forward angles, ensured that the phase space volume 
of the final decaying system was covered efficiently. 

\begin{figure}[t]
\resizebox{0.90\columnwidth}{!}{%
\includegraphics{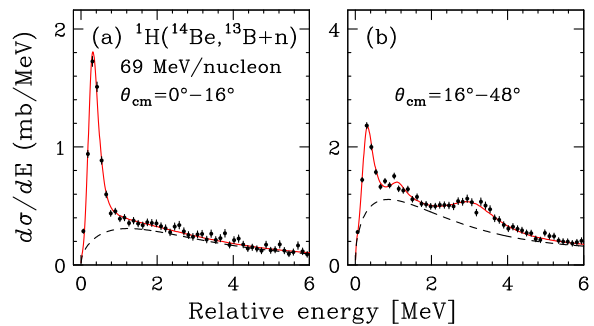}%
}
\caption{Relative energy spectra 
of the $^1\textrm{H}($$^{14}\textrm{Be},$$^{13}\textrm{B}+n)$ reaction 
for (a) $\theta_{\rm cm}$=$0^{\circ}$--$16^{\circ}$ 
and (b) $\theta_{\rm cm}$=$16^{\circ}$--$48^{\circ}$. 
\label{fig:spectrum_fit}}
\end{figure}

Figure~\ref{fig:spectrum_fit} shows relative energy spectra 
obtained at two different angular bins: 
(a) $\theta_{\rm cm}$=$0^{\circ}$--$16^{\circ}$ 
and (b) $\theta_{\rm cm}$=$16^{\circ}$--$48^{\circ}$. 
The effects of finite detector acceptances were corrected. 
Background contributions due to window materials, 
measured with an empty target, 
were less than 0.1 mb/MeV, 
and were subtracted in the figure. 
Error bars are statistical ones. 
A sharp resonance peak, evident at $E_{\rm rel}$$\approx$0.3 MeV, 
is most prominent at forward angles. 
A bump structure at $E_{\rm rel}$$\approx$1.1 MeV 
is visible only at backward angles, 
and might correspond to the reported $4^-$, 
2.08 MeV state observed in the 
$^{14}\textrm{C}$($^7\textrm{Li}$,$^7\textrm{Be}$)$^{14}\textrm{B}$~\cite{Ball73} 
and $^{12}\textrm{C}$($^{14}\textrm{C}$,$^{12}\textrm{N}$)$^{14}\textrm{B}$~\cite{Kalpakchieva00} 
reactions. 
Since this state is weakly populated, 
the energy of 2.08 MeV was adopted in the subsequent analysis. 
Another bump at $E_{\rm rel}$$\approx$3 MeV 
was observed for the first time in this study. 
For the three states no coincident $\gamma$ rays were identified. 

The spectra were analyzed using single Breit-Wigner shape functions 
as described in Ref.~\cite{Ysatou08} 
to extract the resonance energy $E_r$, width ${\it \Gamma}_r$, 
and peak yield. 
The line shape was obtained in a Monte Carlo simulation 
which includes the detector response, beam profile, and target thickness. 
The relative energy resolution was well 
described by the relationship: 
$\Delta E_{\rm rel}$=$0.20\sqrt{E_{\rm rel}}$ MeV(rms). 
For strengths beneath the resonance, 
thermal emission of a neutron from the continuum~\cite{Deak87} 
was assumed. 
A constant background, 
which simulates various effects 
such as overlapping broad resonances and decay to an excited residue 
was also introduced when needed. 
Solid lines in Fig.~\ref{fig:spectrum_fit} 
are the results of the fit; 
the assumed continuum-like strengths are indicated as dashed lines. 
Resulting resonance parameters are summarized in Table~\ref{tbl:resonance_p}. 
The orbital angular momenta ($l$) of decay neutrons 
are also shown. 
\begin{table}[t]
\caption{Resonance parameters in the present analysis. 
The excitation energy of $E_x$=2.08 MeV 
is from Refs.~\cite{Ball73,Kalpakchieva00}.
\label{tbl:resonance_p}} 
\begin{ruledtabular}
\begin{tabular}{dddcc} 
\multicolumn{1}{c}{$E_r$ (MeV)}  & 
\multicolumn{1}{c}
{$E_x$ (MeV)} & 
\multicolumn{1}{c}{${\it \Gamma}_r$ (MeV)}   & $l$ ($\hbar$) & $J^{\pi}$ \\ 
\hline
0.304(4) & 1.27(2)    & 0.16(2)             & 1  & $1^+$    \\
1.11     & 2.08  & \multicolumn{1}{c}{---}  & 2 & $4^-$   \\
3.09(5)  & 4.06(5)    & 
\multicolumn{1}{c}{[1.0(3),1.2(5)]} & \multicolumn{1}{c}{(1,2)} & 
\multicolumn{1}{c}{($3^+$,$3^-$)}  \\
\end{tabular}
\end{ruledtabular}
\end{table}
The excitation energy ($E_x$=$E_r$+$S_n$) of 1.27(2) MeV 
for the peak at $E_{\rm rel}$$\approx$0.3 MeV 
agrees well with the reported value of 1.28(2) MeV 
for the first $1^+$ state in $^{14}\textrm{B}$~\cite{Belbot97,Aoi02}. 
The width of ${\it \Gamma}_r$=0.16(2) MeV 
is consistent with the single-particle estimate of ${\it \Gamma}_{\rm sp}$
=0.31 MeV for an $l$=1 resonance calculated in a Woods-Saxon potential, 
while it is significantly 
larger than ${\it \Gamma}_{\rm sp}$=0.02 MeV for a $d$-wave, 
supporting the assumption of a $p$-wave decay. 
It is to be noted that the single particle width, ${\it \Gamma}_{\rm sp}$, 
is an estimate of the total decay width derived within the one-particle model 
in a simple nuclear potential, and that it provides an upper limit for the 
actual nucleus case where the configuration mixing is generally important. 
For the 2.08 MeV state, 
the width was not reliably extracted due to limited signal-to-background 
ratio near the resonance; 
it was assumed to be dominated by the resolution. 
Note that an experimental upper limit 
of ${\it \Gamma}_r$=0.1 MeV~\cite{Kalpakchieva00} has been reported. 
For the 4.06 MeV state two different decay angular momenta, $l$=1 and 2, 
were considered. 
The deduced widths are shown in Table~\ref{tbl:resonance_p}. 
The resonance energy 
was similar in the two cases and varied within the errors. 

\begin{figure}[t]
\resizebox{0.90\columnwidth}{!}{%
\includegraphics{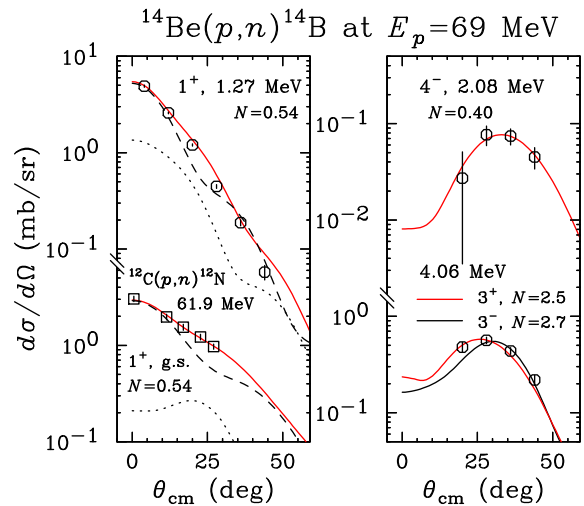}
}
\caption{Angular distributions of the differential cross section 
leading to the three states observed (open circles). 
Data for the $^{12}\textrm{C}(p,n)$$^{12}\textrm{N}(1^+,\textrm{g.s.})$ reaction 
at 61.9 MeV~\cite{Anderson80} (open squares) are also shown. 
Normalization factors applied to the DWBA curves (solid lines) 
are indicated in the figure. 
Additional curves, dashed and dotted lines, 
given for $0^+$$\rightarrow$$1^+$ transitions 
are DWBA results obtained, 
respectively, with only the central and tensor terms of the $NN$ interaction. 
}
\label{fig:neut_14b_ad}
\end{figure}

Figure~\ref{fig:neut_14b_ad} shows 
differential cross sections leading to the observed states. 
These were obtained by subdividing 
the acceptance solid angle to have a spectrum at each angular bin, 
and applying the fitting procedure to extract the peak yields. 
Error bars are statistical ones; 
systematic uncertainties are estimated to be 6\%, 
including those 
in the target thickness, neutron detection efficiency, and fitting procedure. 
The resolution in the scattering angle was estimated 
with the Monte Carlo simulation 
to vary from $\Delta\theta_{\rm cm}$=2.3$^{\circ}$ in rms 
at $\theta_{\rm cm}$=4$^{\circ}$ to 4.5$^{\circ}$ at $\theta_{\rm cm}$=44$^{\circ}$. 

Microscopic DWBA calculations were performed 
to clarify the nature of states observed 
and to have a means to extrapolate the $1^+$ cross section to $q$=0. 
The OMP 
was taken from the semi-microscopic parametrization of JLMB~\cite{Bauge01}. 
Within the local density approximation, 
the local value of the finite nucleus OMP 
was deduced from the nuclear matter OMP. 
The proton and neutron radial density distributions 
were calculated in a mean field 
using the SkX interaction~\cite{Brown98}. 
The effective $NN$ interaction 
was the M3Y interaction~\cite{Bertsch77} 
(force components labeled 1,4,11,14,16,17, and 20 were used). 
The transition density was 
obtained from the shell model using the code \textsc{oxbash}~\cite{OXBASH}. 
The calculations used the WBT Hamiltonian~\cite{Warburton92} 
in the $spsdpf$ model space 
within the 0$\hbar\omega$ configuration. 
The single-particle wave functions were generated in a harmonic oscillator well 
with an oscillator parameter $b$=2.04(6) fm, 
chosen to reproduce the rms matter radius of $^{14}\textrm{Be}$~\cite{Ozawa01}. 
The code \textsc{dw}{\footnotesize 81}~\cite{Raynal70} 
was used for the calculations. 

The DWBA cross sections are shown as solid lines in Fig.~\ref{fig:neut_14b_ad}. 
Normalization factors required for the theoretical curves to reproduce the data 
are also indicated. 
For the calculation of the 1.27 MeV state, 
the wave function of the first $1^+$ shell-model state predicted at 1.01 MeV 
was used, 
while for the 2.08 MeV state 
that of the first $4^-$ state at 1.75 MeV was used. 
The measured cross sections leading to these states, 
which exhibit characteristic angular distribution patterns, 
are well reproduced by the calculations, 
corroborating the earlier spin-parity ($J^{\pi}$) assignments 
of $1^+$ for the 1.27 MeV state~\cite{Aoi02} 
and $4^-$ for the 2.08 MeV state~\cite{Ball73,Kalpakchieva00}.
For the 4.06 MeV state, 
the calculation was repeated by assuming 
different shell-model wave functions with various $J^{\pi}$ values. 
The angular distribution was found to be best described by the curve assuming 
the second $3^+$ state predicted at 4.20 MeV (red line). 
Another possible candidate 
was the second $3^-$ state at 5.70 MeV. 
Although the calculated cross section, 
shown in Fig.~\ref{fig:neut_14b_ad} as a black line, 
is slightly backward peaked, 
it is compatible with the data and 
the possibility could not be completely ruled out. 
With this a tentative $J^{\pi}$ assignment of $3^+$ or $3^-$ 
is made to this state. 
It would be interesting if we could distinguish the two cases 
by experiments with different sensitivity from the present one. 

The extraction of $B(\textrm{GT})$ leading to the $1^+$, 1.27 MeV state 
was done by using Eq.(\ref{eq:proportionality_1}). 
The DWBA framework as described above 
was utilized to calculate the parameter for momentum extrapolation 
and the distortion factor. 
In order to see the contribution from the non-central terms 
of the $NN$ interaction, two kinds of additional DWBA calculations 
were performed: one with only the central terms and the other 
with only the tensor terms, and the results of these calculations are shown 
as dashed and dotted lines in Fig.~\ref{fig:neut_14b_ad}, respectively 
(the contribution from the spin-orbit terms turned out to be negligibly small). 
The ratio of the cross section calculated with only the central terms to that 
with full force components was 0.961 at $\theta$=0$^{\circ}$, 
while it was closer to unity, 1.008, at $q$=0. 
Two additional sets of OMP parameters~\cite{Varner91,Koning03} 
were used to estimate the uncertainty due to the choice of the optical model. 
For calculations of the cross section at $q$=0, 
the distorting potential in the exit channel 
was re-evaluated at a neutron energy 
which satisfies the kinematical constraint. 
The volume integral $|J_{\sigma \tau}|$ of the $NN$ interaction 
can be deduced effectively 
by analyzing the existing $(p,n)$ $0^+$$\rightarrow$$1^+$ transition data 
at a similar incident energy, 
for which the $B(\textrm{GT})$ value is well established. 
The $^{12}\textrm{C}(p,n)$$^{12}\textrm{N}(1^+,\textrm{g.s.})$ reaction data 
at 61.9 MeV~\cite{Anderson80}, 
with the corresponding $B(\textrm{GT})$ value of 0.875(6)~\cite{Ajzenberg90}, 
were used for this purpose. 
In the DWBA analysis, 
three sets of OMP parameters~\cite{Bauge01,Varner91,Koning03} 
and the P(10--16)T shell-model wave function~\cite{Warburton92} were used. 
The oscillator parameter chosen was $b$=1.87 fm~\cite{Comfort81}, 
which reproduces the maxima of the electron scattering form factors. 
A result of the fit of the theoretical curve, 
calculated with the JLMB~\cite{Bauge01} OMP, 
to the data 
is shown as a solid line in Fig.~\ref{fig:neut_14b_ad}. 
Dashed and dotted curves are again DWBA cross sections calculated with only the 
central and tensor terms of the $NN$ interaction, respectively. 
For this transition, switching off the tensor terms had an effect to reduce the 
cross section by 1.3\% ($-$0.8\%) at $\theta$=0$^{\circ}$ ($q$=0). 
By using Eq.(\ref{eq:proportionality_1}), 
$|J_{\sigma \tau}|$ at 61.9 MeV was deduced to be 
212.1$\pm$3.1(stat)$\pm$9.3(syst) MeV\,fm$^3$, 
where the systematic error includes the uncertainty 
due to the choice of optical parameters (2\%). 
The value turns out to be consistent 
with the tabulated ones for the M3Y interaction~\cite{Love80}: 
$|J_{\sigma \tau}|$=211(247) MeV\,fm$^3$ at $E_p$=61(40) MeV. 
From the proportionality between the cross section at $q$=0 
and $B(\textrm{GT})$ [Eq.(\ref{eq:proportionality_1})], 
which assumes $|J_{\sigma \tau}|$ at 69 MeV obtained 
by making energy extrapolation using values for the M3Y interaction, 
the $B(\textrm{GT})$ for the transition 
to the $1^+$ state at 1.27 MeV in $^{14}\textrm{B}$ 
was found to be 0.79$\pm$0.03(stat)$\pm$0.09(syst). 
The systematic error includes the uncertainty 
due to the choice of optical parameters (4\%). 
The value is in good agreement with 
the $\beta$-decay value of $B(\textrm{GT})$=0.80$\pm$0.09~\cite{Aoi02}, 
calculated from the $ft$ value using the relationship: 
$B(\textrm{GT})$=$(g_A/g_V)^{-2}(6147/ft)$~\cite{Harakeh01}. 

The shell-model prediction obtained 
using the WBT interaction~\cite{Warburton92} 
and the effective GT operators~\cite{Chou93}, 
within the 0$\hbar \omega$ basis, 
is $B(\textrm{GT})$=0.88, 
which is consistent with the data. 
This suggests that the ground state of $^{14}\textrm{Be}$ 
is predominantly of 0$\hbar \omega$ nature. 
Precision data and improved knowledge on effective GT operators 
will allow further detailed analyses of the admixtures of higher-$\hbar \omega$ 
intruder components~\cite{Umeya08}, in terms of $B(\textrm{GT})$. 

The $^{14}\textrm{Be}(\textrm{g.s.})$
$\rightarrow$$^{14}\textrm{B}(1^+_1)$ transition 
exhausts only a minor fraction ($\sim$4\%) of the GT sum rule: $3(N$$-$$Z)$=18. 
Major GT strengths are expected to be located 
at previously unexplored 
high excitation energy regions~\cite{Sagawa93,Suzuki97}. 
CE studies probing such strengths are a challenge 
(this might need either multiple-neutron detection 
within the current invariant mass scheme 
or recoiled-neutron detection in the missing mass scheme), 
but will shed new light on our understanding of 
spin-isospin correlations in nuclei 
at extreme conditions of isospin and binding energy. 
The present study, 
quantifying the accuracy for a procedure to extract $B(\textrm{GT})$ 
from forward angle $(p,n)$ cross sections on radioactive beams 
to be of order 15\%, 
paves the way for future investigations in this direction. 

In summary, 
the $(p,n)$ reaction on $^{14}\textrm{Be}$ was measured at 69 MeV/nucleon 
using the invariant mass method. 
In decay energy spectra three resonance states were observed. 
A comparison of measured cross sections with DWBA calculations 
yielded a confirmation of the earlier $J^{\pi}$ assignments 
of $1^+$ and $4^-$ for the 1.27 and 2.08 MeV states, respectively. 
It also led to a tentative $J^{\pi}$ assignment 
of $3^+$ or $3^-$ for a state newly observed at 4.06 MeV. 
By extrapolating the cross section to $q$=0, 
the $B(\textrm{GT})$ leading to the $1^+$ state was deduced, 
which compared well with the $\beta$-decay value, 
providing the first successful demonstration of an experimental capability 
of extracting GT strengths on unstable nuclei 
via decay spectroscopy following CE reactions in inverse kinematics. 

\begin{acknowledgments}
The authors are grateful for the invaluable assistance of the staff 
of RIKEN during the experiment. 
They thank Professor J.A.~Tostevin for reading the manuscript. 
The work was in part supported by the Grant-in-Aid for Scientific Research 
(No.~15740145) of MEXT Japan and the WCU program (R32-2008-000-10155-0) of 
NRF Korea. 
\end{acknowledgments}


\end{document}